\documentclass[10pt, twocolumn, secnumarabic, amssymb, nobibnotes, footinbib, showpacs, aps, pra, superscriptaddress]{revtex4-1}

\usepackage{graphicx}
\usepackage{dcolumn}
\usepackage{amsmath}
\usepackage{natbib}
\usepackage{color}
\usepackage{epstopdf}

\begin{document}

\title{How to observe duality in entanglement of two distinguishable particles?}

\author{Marcin Karczewski}
\affiliation{Faculty of Physics, Adam Mickiewicz University, Umultowska 85, 61-614 Pozna\'n, Poland}
\author{Pawe\l{} Kurzy\'nski}   \email{pawel.kurzynski@amu.edu.pl}  
\affiliation{Faculty of Physics, Adam Mickiewicz University, Umultowska 85, 61-614 Pozna\'n, Poland}
\affiliation{Centre for Quantum Technologies, National University of Singapore, 3 Science Drive 2, 117543 Singapore, Singapore}

\date{\today}


\begin{abstract}
The entanglement between two bosons or fermions can be accessed if there exists an auxiliary degree of freedom which can be used to label and effectively distinguish the two particles. For some types of entanglement between two indistinguishable particles one can observe {\it duality}, i.e., if the entanglement is present in the Hilbert space $\cal{H}$ and  an auxiliary Hilbert space $\cal{H}'$ is used to label the particles, then if we used $\cal{H}$ as a label the entanglement would be present in $\cal{H}'$. For distinguishable particles this effect does not occur because of superselection rules which prevent superpositions of different types of particles. However, it is known that superselection rules can be bypassed if one uses special auxiliary states that are known as reference frames. Here we study properties of reference frames which allow for an observation of a duality in entanglement between two distinguishable particles. Finally, we discuss the consequences of this result from the resource-theoretic point of view.
\end{abstract}



\maketitle


\section{Introduction}

Two polarization-entangled photons can be generated in a standard parametric down-conversion (PDC) experiment. The entanglement can be accessed because the photons can be distinguished by their momentum -- a photon moving left and the one moving right. Interestingly, if the two PDC photons go through polarizing beam splitters and the H-polarized one is sent to Alice, whereas the V-polarized one is sent to Bob, the party can still detect entanglement, but this time it occurs in the momentum degree of freedom. This phenomenon is known as duality in entanglement and does not occur in case of distinguishable particles. 

Duality in entanglement provides an interaction-free test of particle indistinguishability \cite{bose13} and can be observed in various physical implementations \cite{duality1, duality2}.  The idea of the test relies on the fact that for distinguishable particles superselection rules (SSR) \cite{SSR1} restrict the set of all the possible measurements. As a result, the entanglement cannot be observed after the swap of labels. However, such a rule can be lifted by the introduction of a proper additional state, called a reference frame \cite{SSR2,SSR3,SSR4}. In this work we investigate its impact on the duality of entanglement. 

It is well know that in the first quantization picture the symmetrization/anti-symmetrization of the wave function can be sometimes considered as entanglement which cannot be operationally accessed \cite{IE1,IE2,IE3,IE4,IE5,IE6,IE7}. Nevertheless, due to this fact fermionic and bosonic properties can be simulated with distinguishable particles if one prepares them in a proper symmetric or anti-symmetric state. Such a preparation requires entanglement consumption, which in this case can be considered as a resource to simulate identicality of particles. However, it is not obvious that every bosonic/fermionic property can be simulated only by symmetrization/anti-symmetrization. Some properties may also be simulated with alternative resources. 

Here, we ask what alternative resources contained in reference frames can be used to simulate duality in entanglement. In particular, we identify the minimal conditions needed for such a reference frame to enable distinguishable particles to exhibit the duality. The result pinpoints the aspects of indistingiushability captured by the entanglement duality. This highlights issues that should be considered while preparing the duality-based tests of identicity of particles. Moreover, it contributes to a formulation of a resource theory of indistinguishability.

		
\section{Idea}

\begin{figure}[t]
\includegraphics[width=0.62 \textwidth,trim=4 4 4 4,clip]{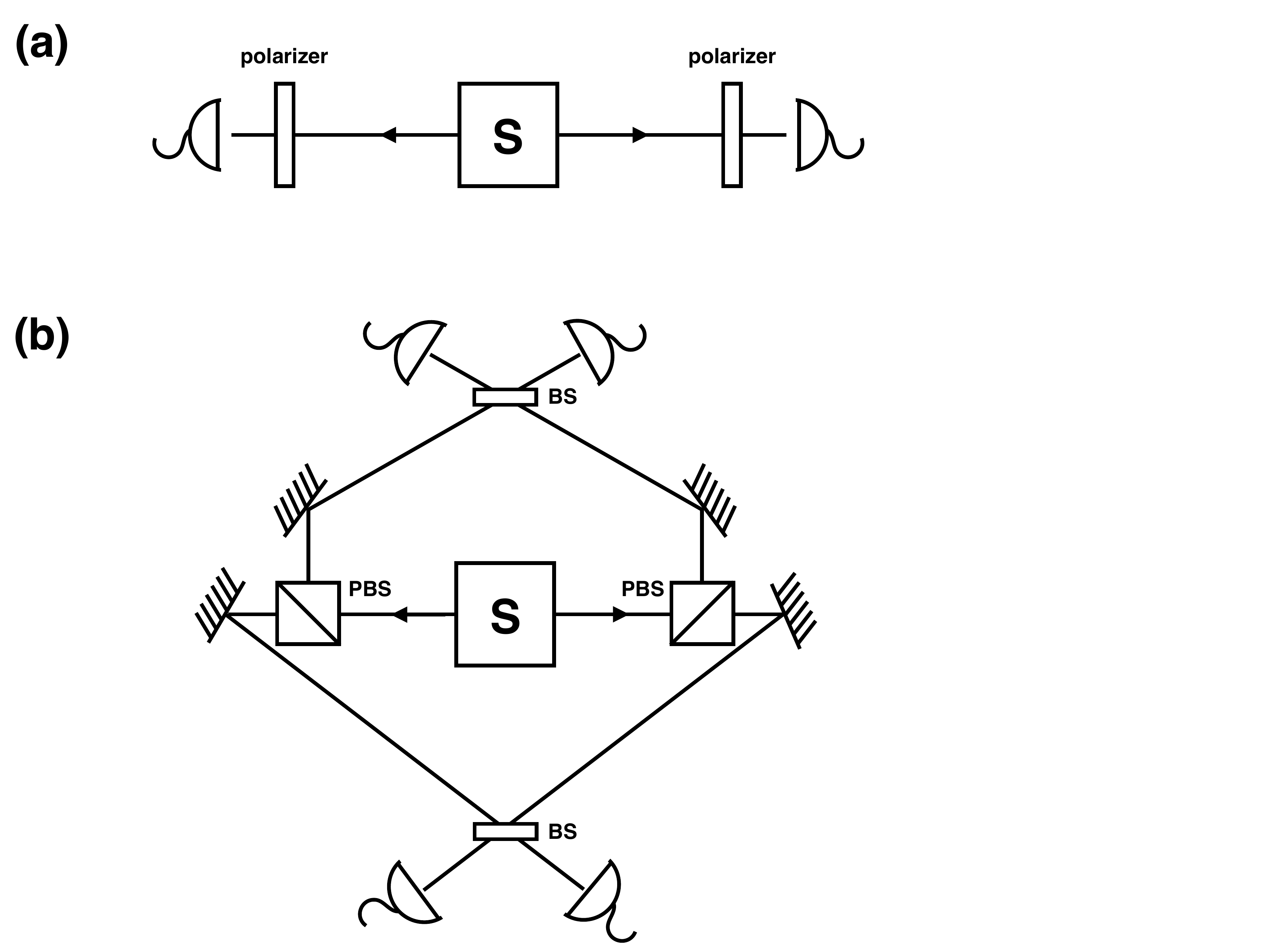}
\caption{ An experimental setup to verify the duality in entanglement. An entangled pair of photons $|\Psi\rangle=|H\rangle_{\bar k}|V\rangle_k+|V\rangle_{\bar k}|H\rangle_k$ is created by the source S. The particles are emitted with opposite  momenta $k$ and $\bar k$ towards Alice and Bob respectively. They can now perform standard polarization measurements to verify the entanglement (a) or use the polarizing beam splitters (PBS) to test its dual form. If they choose the latter (b), the photons are either transmitted or reflected, according to their polarization. As a result, only one particle goes up and one goes down. The beam splitters (BS) can be used to verify entanglement in the momentum degree of freedom.  }
\label{fig: bose}
\end{figure}

We follow Ref. \cite{bose13} and consider two PDC photons 
 \begin{align} \label{eq: bose creation}
\frac{1}{\sqrt2}\big( a_{H,\,k}^\dag  a_{V,\,\bar k}^\dag +  a_{V,\,k}^\dag  a_{H,\,\bar k}^\dag\big)|0\rangle,
 \end{align}
 where $a_{X,\,p}^\dag $ denotes the operator that creates a photon with polarization $X=H,V$ and momentum $p=k,\bar k$.
The above state is clearly entangled if we distinguish particles by their momentum 
\begin{align}
|\Psi\rangle=\frac{1}{\sqrt2}\big( |H\rangle_{k}|V\rangle_{\bar k}+|V\rangle_{k}|H\rangle_{\bar k} \big).
\end{align}
If instead the particles are labeled by the polarization we get
\begin{align}
|\Psi\rangle=\frac{1}{\sqrt2}\big(|k\rangle_{H}|\bar k\rangle_{V} + |\bar k\rangle_{H}|k\rangle_{V}\big).
\end{align}
The fact that the presence of entanglement is independent of labeling is known as duality in entanglement and can be tested in the setup presented in Fig. \ref{fig: bose} (for details see \cite{bose13}).


For a system of distinguishable particles we use different creation operators
 \begin{align} \label{eq: bose creation2}
\frac{1}{\sqrt2}\big( a_{H,\,k}^\dag  b_{V,\,\bar k}^\dag +  a_{V,\,k}^\dag  b_{H,\,\bar k}^\dag\big)|0\rangle.
 \end{align}
 Now the entanglement depends on the choice of indexing. As momentum is consistent with the type of the particle, the state 
 \begin{align}
|\Psi\rangle=\frac{1}{\sqrt2}\big(|H\rangle_{a,\,k}|V\rangle_{b,\,\bar k}+|V\rangle_{a,\,k}|H\rangle_{b,\,\bar k}\big).
\end{align}
is entangled. On the other hand, labeling by polarization leads to
\begin{align}
|\Psi\rangle=\frac{1}{\sqrt2}\big(|k,\,a\rangle_{H}|\bar k,\,b\rangle_{V} + |\bar k,\, b\rangle_{H}|k,\,a\rangle_{V}\big). 
\end{align}
Such a state is operationally mixed, as its entanglement cannot be observed. The standard measurements can only check if the momentum state was $|k\rangle$ or $|\bar{k}\rangle$, but in order to detect entanglement one also needs to measure some other complementary states, say $\alpha |k\rangle + \beta |\bar{k}\rangle$. However, the second measurement can only be done if $\alpha \beta=0$. This follows from SSR which prohibit states superposing different particle types from being the eigenstates of quantum observables and the fact that $|k\rangle$ is associated with the particle $a$ and $|\bar{k}\rangle$ with the particle $b$. Because of the distinguishability of particles, there is no duality in entanglement for this system.

In the first quantization picture the state (\ref{eq: bose creation}) can be written as
\begin{align}
|\Psi\rangle = & \frac{1}{2} \big( |H,k\rangle_1 |V,{\bar k}\rangle_2+|V,k\rangle_1 |H,{\bar k}\rangle_2 \nonumber \\  + &  |H,\bar k\rangle_1 |V,{k}\rangle_2+|V,\bar k\rangle_1 |H,{ k}\rangle_2 \big),
\end{align}
where the particles are labeled as 1 and 2, though they are indistinguishable. This state can be mathematically rewritten as a hyper-entangled state
\begin{align}\label{hyper}
|\Psi\rangle = \frac{1}{2}\big( |H\rangle_1 |V\rangle_2+|V\rangle_1 |H\rangle_2 )\otimes (|k\rangle_1 |{\bar k}\rangle_2+|\bar k\rangle_1 |{k}\rangle_2 \big),
\end{align}
although one needs to remember that only one type of entanglement can be accessed because the other one needs to take care of state symmetrization. However, if the two particles were distinguishable, the state (\ref{hyper}) would be operationally hyper-entangled and it would pass the indistinguishability test based on duality in entanglement. In the second quantization picture it is of the form 
\begin{align} 
\frac{1}{2}\big( a_{H,\,k}^\dag  b_{V,\,\bar k}^\dag +  a_{V,\,k}^\dag  b_{H,\,\bar k}^\dag + a_{V,\bar k}^\dag  b_{H, k}^\dag + a_{H,\bar k}^\dag  b_{V, k}^\dag \big)|0\rangle.
 \end{align}
One can interpret this in the following way -- by adding entanglement to the system the state (\ref{eq: bose creation2}) becomes symmetric and behaves like a state of indistinguishable particles. 

Now, we ask if the state (\ref{eq: bose creation2}) can pass the duality test without symmetrization. This is possible if the measurements are performed on an extended system $\rho=\rho_S\otimes\rho_A$. Here, $\rho_S=|\Psi\rangle\langle \Psi|$ is the state of the original system and $\rho_A$ is the state of an ancilla, which is commonly known as a reference frame. To illustrate the idea we  consider  $\rho_A=\rho_S$. This way the extended system is in the state $\rho=\big(|\Psi\rangle\otimes|\Psi\rangle\big)\big(\langle\Psi|\otimes\langle\Psi|\big)$, where
  \begin{align}
|\Psi\rangle\otimes|\Psi\rangle=\frac{1}{2}\big( a_{H,\,k}^\dag  b_{V,\,\bar k}^\dag +  a_{V,\,k}^\dag  b_{H,\,\bar k}^\dag\big)^{\otimes2}\,|0\rangle\otimes|0\rangle
 \end{align}
 and the tensor product denotes the fact that each copy of the system occupies a different mode.
 
For convenience we set
\begin{align}
|k,\,a\rangle &\equiv |0\rangle, \label{n1} \\
|\bar k,\,b\rangle&\equiv |1\rangle, \label{n2}
\end{align}
so that the state (6) becomes
\begin{align}\label{simplestate}
|\Psi\rangle=\frac{1}{\sqrt 2} (|0\rangle |1\rangle + |1\rangle |0\rangle).
\end{align}
In the above we skipped the subindices $H$ and $V$ and use the convention that the first state in the product is $H$-polarized and the second $V$-polarized. We stress that this is just a rewriting and the above state is still operationally mixed due to the underlying SSR, i.e., observing the state $\alpha |0\rangle + \beta |1\rangle$ is only possible if $\alpha \beta =0$. However, if we add its copy we obtain
\begin{align}\label{2copies}
|\Psi\rangle\otimes|\Psi\rangle=\frac{1}{2}(\underbrace{|00\rangle |11\rangle +|11\rangle |00\rangle}_{\text{operationally mixed part}} +\underbrace{|01\rangle |10\rangle +|10\rangle |01\rangle}_{\text{entangled part}} ),
\end{align}
where we used the notation separating the subsystems of Alice and Bob, i.e., $|00\rangle|11\rangle$ means that Alice has a state $|00\rangle$ and Bob has $|11\rangle$. The state (\ref{2copies}) consists of the operationally mixed and entangled parts. This is because $\alpha |00\rangle + \beta |11\rangle$ is only possible if $\alpha \beta =0$, but $\alpha |01\rangle + \beta |10\rangle$ is allowed for any $\alpha$ and $\beta$ (up to the normalisation constraint $|\alpha|^2+|\beta|^2=1$).

In order to verify the entanglement in the above state we consider the Peres-Horodecki criterion \cite{Peres,Horodecki}, i.e., the negativity of the partially transposed density matrix. However, the matrix to which we apply this condition needs to be modified. Because some coherences are unobservable due to the SSR, they need to be excluded from the effective density matrix (we locally apply the so-called {\it twirling} operation \cite{SSR3,SSR4})
\begin{eqnarray}
\rho_{eff}  = &\frac{1}{4}& (|00\rangle\langle 00||11\rangle\langle 11| + |11\rangle\langle 11||00\rangle\langle 00| \nonumber  \\ 
&+&  |01\rangle\langle 01||10\rangle\langle 10| + |10\rangle\langle 10||01\rangle\langle 01|  \nonumber \\ 
&+& |01\rangle\langle 10||10\rangle\langle 01|  + |10\rangle\langle 01||01\rangle\langle 10| ).
\end{eqnarray}
The first four terms are diagonal, whereas the last two correspond to observable coherencies and are responsible for the entanglement. After applying a partial transposition we obtain
\begin{eqnarray}
\rho_{eff}^{\Gamma}  = &\frac{1}{4}& (|00\rangle\langle 00||11\rangle\langle 11| + |11\rangle\langle 11||00\rangle\langle 00| \nonumber  \\ 
&+&  |01\rangle\langle 01||10\rangle\langle 10| + |10\rangle\langle 10||01\rangle\langle 01|  \nonumber \\ 
&+& |01\rangle\langle 10||01\rangle\langle 10|  + |10\rangle\langle 01||10\rangle\langle 01| ).
\end{eqnarray}
This is a block diagonal matrix and the last block, corresponding to the last two terms, has eigenvalues $\pm\frac{1}{4}$. This confirms that the state is entangled. 

The example shows an idea of bypassing SSR with an additional subsystem. However, it is not clear that the observable entanglement in the effective state originates from the original subsystem, since the additional one is also entangled \cite{SSR4}. We will resolve this issue in the next section.


\section{Separable reference frame}

Let us consider a reference frame in a Werner state \cite{Werner}
\begin{align}\label{sr}
\rho_{A} = \frac{1-p}{4}\openone+p|\Psi\rangle\langle\Psi|,
\end{align}
where $0 \leq p \leq 1$, $|\Psi\rangle$ is the same as in (\ref{simplestate}) and the identity is expressed in terms of states (\ref{n1}) and (\ref{n2})
\begin{align}
\openone=|0\rangle\langle 0||0\rangle\langle 0| + |0\rangle\langle 0||1\rangle\langle 1| + |1\rangle\langle 1||0\rangle\langle 0| + |1\rangle\langle 1||1\rangle\langle 1|.
\end{align}
The state (\ref{sr}) is separable for $p < \frac{1}{3}$ \cite{Werner}. 

The total state of the system is
\begin{align}\label{rp}
\rho_p=\rho_S\otimes\rho_A=\frac{1-p}{4} |\Psi\rangle\langle\Psi|\otimes \openone+p|\Psi\rangle\langle\Psi|\otimes|\Psi\rangle\langle\Psi|.
\end{align}
Coherencies in the first term are unobservable due to the SSR, therefore this part of the state is effectively diagonal. The second term is the same as in the example from the previous section. 

In order to confirm the entanglement in the above state we apply once again the Peres-Horodecki criterion. The effective partially transposed density matrix is block diagonal and the only relevant block is a $2\times 2$ submatrix, which contains an off-diagonal term and is proportional to the one considered before
\begin{align}
\frac{p}{4}( |01\rangle\langle 10||01\rangle\langle 10|  + |10\rangle\langle 01||10\rangle\langle 01| ).
\end{align}
Its eigenvalues are $\pm \frac{p}{4}$ and we see that the total state $\rho_p$ is entangled for any value of $p>0$. In particular, if $0<p<\frac{1}{3}$ we can confirm entanglement in momentum and at the same time we know that the state of the reference frame (\ref{sr}) is separable. Therefore, the entanglement comes solely from the original state and the reference frame is only used to activate it.


\section{Discussion}
	
The above examples show that it is not necessary that the reference frame is entangled, but rather that it contains non-zero off-diagonal terms. In addition, the reference frame has two important additional features. Just like in the case of particle number SSR, although the state $\rho_A$ is separable, it cannot be prepared locally \cite{SSR4}. This is because the local preparation would require a violation of local SSR. In addition, although particles $a$ and $b$ in the original system are different, preparation of a reference frame capable of activating the dual entanglement requires the same type of particles (either bosons or fermions). If the reference frame consisted of other particles, say $c$ and $d$, one would not be able to bypass the SSR.   

One may ask what type of physical property of the reference frame (what type of a resource) is responsible for the activation of the dual form of entanglement. To answer this question we first recall some previous results on nonclassical correlations in the presence of SSR. It was shown in \cite{SSR4} that in the presence of a particle number SSR in order to violate Bell inequality with the entanglement contained solely in the original system one needs a reference frame with a non-zero {\it superselection-induced variance} (SIV) \cite{SIV}. SIV is a resource that arises in bipartite systems which are subject to the particle number SSR. It corresponds to a local uncertainty of the particle number, despite the fact that the global particle number is fixed. It is defined as the variance of the local particle number
\begin{align}
V(\psi) = 4\left(\langle \psi| N_A^2\otimes\openone |\psi\rangle - \langle \psi| N_A\otimes\openone  |\psi\rangle^2\right),
\end{align}
where the factor of four is due to normalisation \cite{SIV}. For mixed states one can introduce SIV of formation, which is analogous to the entanglement of formation (EOF) \cite{EOF}
\begin{align}
V_F^{SSR}(\rho) = \min_{\{p_i,\psi_i\}}\sum_i p_i V(\psi_i),
\end{align}
where $p_i$ represents a probability distribution over pure states $|\psi_i\rangle$ that create the mixed state $\rho$. However, contrary to EOF, the minimization is not over all possible pure states, but only over those obeying SSR.   

In our case the SSR corresponds to a lack of a superposition between different particle types. This is equivalent to saying that every pure state needs to have a well defined global number of particles $a$ and $b$. We recall the definitions (\ref{n1}) and (\ref{n2}) where $|0\rangle$ was associated with particle $a$ and $|1\rangle$ was associated with particle $b$. The ability to activate dual entanglement comes from the off-diagonal terms in the reference frame. They originate from a superposition $|0\rangle|1\rangle + |1\rangle|0\rangle$ which is the only element of the state $\rho_A$ (\ref{sr}) for which the local numbers of $a$ and $b$ are uncertain. One may therefore associate this uncertainty with the resource that is responsible for the activation of the dual entanglement. It can be measured by the SIV of one type of particle, say particle $a$, which we will denote as SIV$_a$.  Following the result in \cite{SIV2} we have
\begin{align}
V_{F}^{SSR}(\rho_A)_a = \frac{p^2}{2(1+p)}.
\end{align}
If we require that $\rho_A$ is separable, then SIV$_a$ must be less than $\frac{1}{24}$.

To conclude, we argued that the variance of the local particle number (SIV$_a$) is a resource which can activate dual entanglement in an entangled state of distinguishable particles. This shows that some properties of indistinguishable particles can be simulated with distinguishable ones without the need of state symmetrization/anti-symmetrization, provided one has an access to a properly engineered reference frame. It would be interesting to show that other features, that are commonly considered to be typical bosonic or fermionic properties (such as bunching or anti-bunching \cite{HOM}), can also be simulated with SIV$_a$ or some different resource other than symmetrization/anti-symmetrization, which in most cases requires entanglement.
	
{\it Acknowledgements.} This work is supported by the National Science Centre in Poland through the NCN Grant No. 2014/14/E/ST2/00585.



\end{document}